\documentclass[aps,pra,showpacs,twocolumn,amsmath,amssymb]{revtex4}
\usepackage{graphicx}
\usepackage{amsmath}
\usepackage{bigstrut}
\usepackage{color}
\usepackage{dcolumn}
\usepackage{bm}
\usepackage{wasysym}
\usepackage{braket}

\newcommand{\etal}{\textit{et al.}}
\newcommand{\Xstate}{\mbox{$X^1\Sigma^+$}}
\newcommand{\astate}{\mbox{$a^3\Sigma^+$}}

\begin{document}

\title{Feshbach spectroscopy and scattering properties of ultracold Li+Na mixtures}

\author{T. Schuster,$^1$\footnote{E-Mail: naliFR@matterwave.de} R. Scelle,$^1$ A. Trautmann,$^1$ S. Knoop,$^{1,2}$ M. K. Oberthaler,$^1$ M. M. Haverhals,$^3$ M.\,R.\,Goosen,$^{3}$ S.\,J.\,J.\,M.\,F.\,Kokkelmans,$^3$ E. Tiemann$^4$}

\affiliation{$^{1}$Kirchhoff-Institut f\"{u}r Physik, Universit\"{a}t Heidelberg, Im Neuenheimer Feld 227, 69120 Heidelberg, Germany\\
$^{2}$LaserLaB, VU University Amsterdam, De Boelelaan 1081, 1081 HV Amsterdam, The Netherlands\\
$^{3}$Eindhoven University of Technology, P. O. Box 513, 5600 MB Eindhoven, The Netherlands\\
$^{4}$Institute of Quantum Optics, Leibniz Universit\"{a}t Hannover, 30167 Hannover, Germany}

\date{\today}

\begin{abstract}
We have observed 26 interspecies Feshbach resonances at fields up to 2050\,G in ultracold $^6$Li+$^{23}$Na mixtures for different spin-state combinations. Applying the asymptotic bound-state model to assign the resonances, we have found that most resonances have $d$-wave character. This analysis serves as guidance for a coupled-channel calculation, which uses modified interaction potentials to describe the positions of the Feshbach resonances well within the experimental uncertainty and to calculate their widths. The scattering length derived from the improved interaction potentials is experimentally confirmed and deviates from previously reported values in sign and magnitude. We give prospects for $^7$Li+$^{23}$Na and predict broad Feshbach resonances suitable for tuning.
\end{abstract}


\maketitle

\section{Introduction\label{Introduction}}

Magnetically tunable Feshbach resonances \cite{tiesinga1993tar,chin2010fri} play an important role in ultracold atomic gases. Their presence allows tunable interaction strength and coherent association of ultracold molecules. For ultracold atomic mixtures, interspecies Feshbach resonances are used in the creation of ultracold polar molecules \cite{ni2008ahp} and are required to observe heteronuclear Efimov trimer states \cite{barontini2009ooh}. They also allow the investigation of many interesting many-body phenomena in ultracold mixtures \cite{ospelkaus2006toh,best2009roi,trenkwalder2011heo,will2011cio}. Since their first observation \cite{stan2004oof,inouye2004ooh}, interspecies Feshbach resonances have been studied in many experiments. Feshbach resonances in Bose-Fermi mixtures give rise to a rich palette of physical phenomena \cite{bortolotti2008gmf,marchetti2008psa,fatini2010pac,ludwig2011qpt,yu2011sco}.

The Bose-Fermi mixture of $\textsuperscript 6$Li+Na \cite{hadzibabic2002tsm,hadzibabic2003fii} has so far mainly been used to efficiently prepare a degenerate Fermi gas of $\textsuperscript 6$Li. In combination with a Bose-Einstein condensate (BEC) of Na, this mixture is a candidate to show boson mediated interactions between fermions \cite{bijlsma2000pei,heiselberg2000fii}. Interspecies Feshbach resonances have not been utilized in $\textsuperscript 6$Li+Na, despite of the early experimental observation of three resonances \cite{stan2004oof}. 

We have created an ultracold $\textsuperscript 6$Li+Na mixture in different spin-state combinations and searched for interspecies Feshbach resonances at magnetic fields of up to 2050\,G. The observed 26 interspecies Feshbach resonances have been assigned using the asymptotic bound-state model (ABM) \cite{wille2008eau,tiecke2010abs}. Most of the resonances found are of $d$-wave character, i.\,e.~caused by molecular states with rotational angular momentum $l$=2, coupled via the weak magnetic dipole interaction to the $s$-wave continuum. The observed s-wave resonances give rise to very weak loss features, even weaker than some of the $d$-wave resonances, in agreement with a coupled-channel calculation based on the full interaction potentials. We show that the previous assignment of Ref.\,\cite{stan2004oof} is incorrect, and as a result also the predictions of Feshbach resonances and scattering lengths in Ref.\,\cite{gacesa2008fri}.

\begin{figure}[b]
\includegraphics[width=8.5cm]{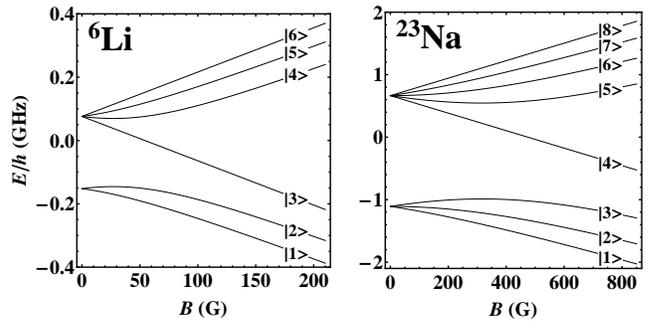}
\caption{The magnetic field dependence of the atomic ground state energies of $\textsuperscript 6$Li and $^{23}$Na. The indicated labeling of the different Zeeman levels is used throughout the paper.\label{Zeeman}}
\end{figure}

This paper is organized as follows. In Sec.\,\ref{Exp} we outline our experimental procedure for the preparation of an ultracold $\textsuperscript 6$Li+Na mixture (Sec.\,\ref{prep}), the Feshbach spectroscopy measurements (Sec.\,\ref{Fspec}), and additional experiments providing information about sign and magnitude of the interspecies singlet and triplet scattering lengths (Sec.\,\ref{Add}). In Sec.\,\ref{Theory} we give an assignment of the observed resonances using ABM (Sec.\,\ref{ABM}). We compare the observed resonance positions with a coupled-channel calculation (Sec.\,\ref{CC}) using modified interaction potentials.  This results in the determination of the resonance widths and background scattering lengths, latter in good agreement with the experimental observations. In Sec.\,\ref{7LiNa} we apply these potentials to the isotopologue $^7$Li+Na and give prospects of possible Feshbach resonances in this Bose-Bose mixture. We conclude and give an outlook in Sec.\,\ref{concl}. Throughout the paper we use a labeling of the Zeeman levels in $\textsuperscript 6$Li and Na as indicated in Fig.\,\ref{Zeeman}.

\section{Experiment}\label{Exp}

\subsection{Preparation of ultracold samples}\label{prep}

Our experimental procedure to obtain an ultracold $\textsuperscript 6$Li+Na mixture is based on the strategy of Ref.\,\cite{hadzibabic2003fii}. In short, we start with a double species magneto-optical trap and load the atoms into a cloverleaf Ioffe-Pritchard magnetic trap, after optically spin-polarizing both Li and Na. Purification of Na with a microwave (MW) sweep at an offset magnetic field results in a stable Na~$|8\rangle$ and Li~$|6\rangle$ mixture. After forced evaporative cooling of Na using the $|8\rangle$$\rightarrow$$|1\rangle$ transition and simultaneous sympathetic cooling of Li, we load the sample into a crossed beam optical dipole trap (ODT) at $\lambda=1064$\,nm. The geometric mean trap frequencies are $\bar{\omega}/2\pi=102(1)$\,Hz and 218(1)\,Hz for Na and Li, respectively. With typically $10^6$~Na and $10^5$~Li atoms and a temperature of 1\,$\mu$K, we get $T/T_C \approx 2$ for Na and $T/T_F \approx 1$ for Li.

We prepare Na in state $|1\rangle$ by a rapid adiabatic passage (RAP) on the $|8\rangle$$\rightarrow$$|1\rangle$ MW transition sweeping the magnetic field around 1\,G. For the preparation of the states $|2\rangle$ or $|3\rangle$ we apply a second RAP on the $|1\rangle$$\rightarrow$$|2\rangle$ and $|2\rangle$$\rightarrow$$|3\rangle$ radiofrequency (RF) transition with a 20\,G offset field. A similar scheme is applied for Li. Specifically, for preparation of the $|3\rangle$ state we omit the $\ket{6}\rightarrow\ket{1}$ MW RAP and directly do the RF transfer $|6\rangle$$\rightarrow$$|5\rangle$$\rightarrow$$|4\rangle$$\rightarrow$$|3\rangle$. It is important to note that in all spin channels the lifetime of the resulting mixtures is longer than 10\,s.

\subsection{Feshbach Spectroscopy}\label{Fspec}

For the Feshbach spectroscopy, we ramp the magnetic field a few Gauss below the resonance, where the system is allowed to thermalize for 0.5\,s. Subsequently, we switch the magnetic field to the desired value, wait for a hold time ranging from 20\,ms to 2\,s such that on resonance a detectable number of Li atoms is still left, which prevents systematic errors due to saturation effects. Na and Li are imaged for different time of flights after ramping down to zero magnetic field. For applying magnetic fields below 1200\,G, the antibias coils of the cloverleaf trap, which are in Helmholtz configuration, are sufficient. Higher magnetic fields are reached adding the pinch coils of the cloverleaf trap with a magnetic field inhomogeneity of at most 10\,mG over the sample, not being a limitation to our Feshbach spectroscopy.

\begin{figure}[b]
\includegraphics[width=8.5cm]{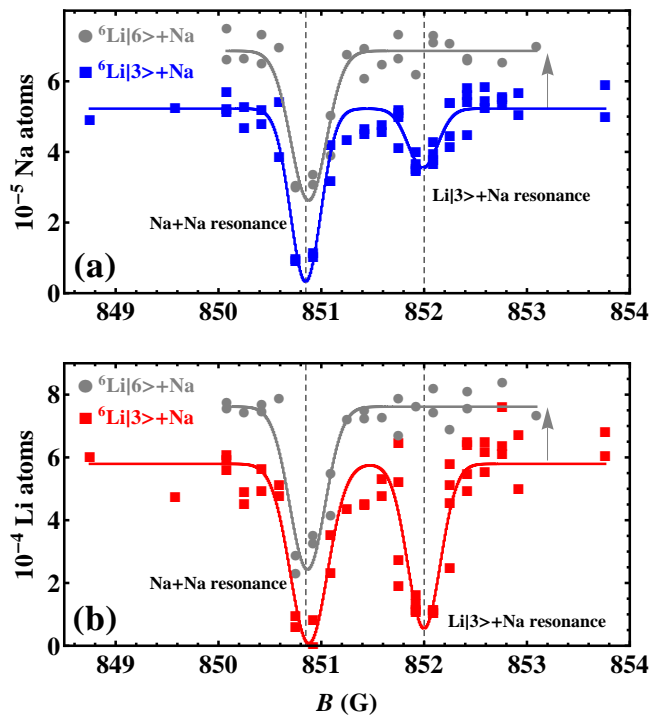}
\caption{(color online) Measured $\textsuperscript 6$Li+Na trap loss spectrum around 850\,G, showing the remaining Na (a) and $\textsuperscript 6$Li (b) atom numbers after a hold time of 1\,s. The solid lines are Gaussian fits. Three-body loss associated with the NaNa resonance in $\ket{1}$ at 851\,G \cite{knoop2011fsa} can involve a Na or a Li atom as third partner and thus leads to loss of Li atoms at resonance, independent of the Li spin channel (gray symbols for Li $|6\rangle$ shifted by an offset for clarity). The loss feature at 852\,G exclusively appears when Li is in the $|3\rangle$ state and can thus be identified as interspecies resonance.\label{FRexp}}
\end{figure}

We take the magnetic field value of maximal loss of Li atoms as the resonance position $B_0^{\rm exp}$, which we obtain by fitting Gaussian lineshapes to our Li loss curves. In magnetic field regions with nearby intraspecies resonances, we carefully checked each loss feature in the mixture by repeating the Feshbach spectroscopy with a single species sample or a different spin-state mixture, an example of which is shown in Fig.\,\ref{FRexp}. To assign the observed loss features as LiNa Feshbach resonances, knowledge of the intraspecies Feshbach resonances is important. For Na \cite{inouye1998oof,stenger1999sei,knoop2011fsa} the Feshbach spectrum is well studied \cite{NaNaFR}, and for $\textsuperscript 6$Li only the $p$-waves at 159\,G and 215\,G for the $|1\rangle$ and $|2\rangle$ states \cite{zhang2004pwf,schunck2005fri}, respectively, have to be considered. 

\begin{table}
\caption{Overview of the experimentally and theoretically obtained results on Feshbach resonances, sorted by quantum number $|M_F|$. In column ``Exp.'' the position of maximum loss $B_0^{\rm exp}$ is reported, which is determined by Gaussian fits to the loss features. As error we give the rms width of the RF calibration signal, a value covering the range from 0.2\,G to 0.9\,G due to the different conditions under which the data were obtained. The other columns show the Feshbach resonance positions from the ABM fit, and the positions and widths $\Delta$ from coupled-channel calculation. The theoretical positions are given by their deviation from the experimental value, i.\,e.~$\delta B_0^{\rm ABM}\equiv B_0^{\rm exp}-B_0^{\rm ABM}$ and $\delta B_0^{\rm CC}\equiv B_0^{\rm exp}-B_0^{\rm CC}$. For the two resonances of asymptote $|6\rangle$+$|1\rangle$ no widths are given (see Sec.~\ref{CC}). The assignment of each resonance in terms of $s$- or $d$-wave resonance is given in the last column ($s$-waves in bold).}
\label{FRlist}

\begin{ruledtabular}
\begin{tabular}{c  c  r  c  c c  c }
         & & Exp.             & ABM                       & \multicolumn{2}{c}{Coupled-Channel}                 &   \\[5pt]
$\textsuperscript 6$Li+Na & $M_F$ & $B_0^{\rm exp}$(G) & $\delta B_0^{\rm ABM}$(G) & $\delta B_0^{\rm CC}$(G) & $\Delta$(mG)  &  \\[1pt]
\hline\\[-8pt]
$|2\rangle$+$|1\rangle$ & 1/2 &  771.8(5)              & \!\!-0.7   & \!\!-0.190    & 10   & $d$   \\
                        &     &  822.9(5)              &  0.5       &  0.050        & 0.5  & $d$   \\
                        &     & 1596.8(4)              &  0.2       & \!\!-0.314    & 5    & $d$  \\
                        &     & 1716.7(3)              &  0.0       &  0.231        & 0.2  & $d$  \\
$|1\rangle$+$|3\rangle$ &-1/2 & 1002.3(5)              & \!\!-0.6   & \!\!-0.209    & 9    & $d$   \\
                        &     & 1088.5(5)              &  0.2       & \!\!-0.301    & 0.9  & $d$  \\
$|3\rangle$+$|1\rangle$ &-1/2 &  800.9(2)              & \!\!-0.4   &  0.096        & 10   & $d$   \\
                        &     &  852.0(7)              &  0.2       & \!\!-0.271    & 0.5  & $d$   \\
                        &     & $\mathbf{1566.3(8)}$   &  0.1       &  0.023        & 0.03 & $\mathbf{s}$  \\
                        &     & 1597.5(7)              &  0.4       & \!\!-0.144    & 6    & $d$  \\
                        &     & 1717.3(2)              & \!\!-0.2   &  0.038        & 0.3  & $d$  \\
\hline\\[-8pt]
$|1\rangle$+$|1\rangle$ & 3/2 & 745.2(3)               & \!\!-0.3   &  0.175        & 10   &   $d$   \\
                        &     & 759.0(3)               &  0.5       &  0.022        & 0.02 & $d$ \\
                        &     & 795.2(2)               &  0.5       & \!\!-0.020    & 0.5  & $d$ \\
                        &     & $\mathbf{1510.4(3)}$   &  0.0       & \!\!-0.024    & 0.04 & $\mathbf{s}$ \\
                        &     & 1596.5(5)              &  0.5       &  0.009        & 5    & $d$  \\
                        &     & 1715.6(8)              & \!\!-0.3   &  0.034        & 0.3  & $d$  \\
                        &     & 1908.9(7)              &  0.4       & \!\!-0.350    & 0.04 & $d$  \\
                        &     & 2046.9(9)              &  0.5       & \!\!-0.608    & 4    & $d$  \\
$|2\rangle$+$|3\rangle$ &-3/2 & 1031.7(3)              & \!\!-0.3   &  0.166        & 9    & $d$  \\
                        &     & 1117.3(6)              &  0.0       & \!\!-0.511    & 0.8  & $d$  \\
                        &     & 1902.4(6)              & \!\!-0.3   & \!\!-0.045    & 0.1  & $d$   \\
$|3\rangle$+$|2\rangle$ &-3/2 &  913.2(6)              & \!\!-0.3   &  0.108        & 9    & $d$   \\
                        &     & $\mathbf{1720.5(3)}$   & 0.0        & \!\!-0.103        & 0.06 & $\mathbf{s}$   \\
\hline\\[-8pt]
$|6\rangle$+$|1\rangle$ & 5/2 & 1575.8(9)              &  0.9       & \!\!-0.014    & ---   & $d$\\
                        &     & 1700.4(7)              &  1.5       & \!\!-0.040    & ---   & $d$ \\

\end{tabular}
\end{ruledtabular}
\end{table}

The magnetic field is calibrated for each Feshbach resonance on the respective position of maximal Li loss by means of RF spectroscopy on the Na $|1\rangle$$\rightarrow$$|2\rangle$ transition. The rms width of the RF spectroscopy is taken as statistical uncertainty of the resonance position $B_0^{\rm exp}$. Since the widths of the observed Li loss features are comparable to our experimentally given magnetic field fluctuations, a deduction of the Feshbach resonance widths $\Delta$ is not possible with our data.

The results of the Feshbach spectroscopy are summarized in Table\,\ref{FRlist}. We have observed 26 Feshbach resonances at magnetic fields of up to 2050\,G for different hyperfine substate combinations, 23 of which had not yet been observed before. The three resonances at low fields in the lowest channel $|1\rangle$+$|1\rangle$ have been previously reported by Stan\,\etal\,\cite{stan2004oof} with resonance positions 746.0(4), 759.6(2) and 795.6(2)\,G, which are at slightly higher values.

\subsection{Experimental determination of scattering lengths}\label{Add}

Additionally to the determination of the Feshbach spectrum, we perform measurements to obtain information about the sign and magnitude of the interspecies scattering length $a$. This will later serve as guidance for the ABM resonance assignment by restricting the $s$-wave bound state energy. 

To determine the absolute value of the triplet scattering length $|a_t|$ experimentally, we excite the dipole mode in the ODT with high amplitude $x_0=30\,\mu$m. In the subsequent oscillation, the sodium and lithium clouds both having a $1/e^2$-radius of $\sigma_x=18\,\mu$m penetrate each other at the frequency $2\omega_{\rm Li}^x$. After an interspecies scattering event, the colliding atoms do not contribute to the coherent oscillation signal any more. With sodium being in the majority as described above, the amplitude of the lithium oscillation decays as
\begin{equation}
	\Gamma =  \sigma\, 2\omega_{\rm Li}^x \int \mathrm dV~n_{\rm Na}(x,y,z)\, \frac{n_{\rm Li}^{(2)}(y,z)}{N_{\rm Li}} \,.
\end{equation}
Here, $\sigma = 4\pi a^2$ is the interspecies elastic collision cross section, $N_{\rm Li}$ the total number of Li atoms,   $n_{\rm Na}(x,y,z)$ the sodium density composed of BEC and thermal cloud and $n_{\rm Li}^{(2)}(y,z)=\int\! \mathrm dx~n_{\rm Li}(x,y,z)$ the lithium density integrated along the direction of oscillation. From a mixture of atoms in the spin-stretched state Li$|6\rangle$+Na$|8\rangle$, where scattering is restricted to the triplet potential, we obtain $|a_t|=69(13)a_0$ \cite{schuster2012phd}.

The sign of the scattering lenght can be infered by comparing the in-situ density profiles of an ultracold lithium cloud with and without a sodium background. For that purpose, we remove the sodium atoms with a resonant light pulse after evaporative cooling of the mixture in the ODT. As can be calculated with the absolute value of the scattering length obtained above, this process leads to a lithium atom loss of less than 1\%, a value below our typical atom number fluctuation. The consecutively observed broadening of the lithium density profile clearly indicates an attractive interspecies interaction $a<0$. 

An insight into the difference between singlet and triplet scattering length $a_s$ and $a_t$ is provided by measuring the two-body loss rate due to spin-exchange collisions. These processes induce changes of the magnetic quantum numbers $m_f^{\rm Na}$ and $m_f^{\rm Li}$ of the two colliding atoms while leaving the total $M_F=m_f^{\rm Na}+m_f^{\rm Li}$ constant. Using Fermi's golden rule, these processes occur at a rate \cite{pethick2004}
\begin{equation}
	\label{eq:diff}
	P_{i \rightarrow f} = 4\pi\sqrt{2E/\mu}\,(a_s-a_t)^2 \left|\braket{f|\vec{s}_1\cdot\vec{s}_2|i}\right|^2\,.
\end{equation}
Here, $\mu$ denotes the reduced mass and $E$ the energy difference between initial state $\ket{i}$ and final state $\ket{f}$. The term $\eta\equiv\braket{f|\vec{s}_1\cdot\vec{s}_2|i}$, with $\vec{s}_{1,2}$ being the electron spin operator of atom 1,2, induces the spin exchange, i.\,e. lowers the spin of atom 1~while raising the spin of atom~2 and vice versa. Eq.\,(\ref{eq:diff}) is exact in the limit of vanishing coupling strength of the magnetic dipole and hyperfine interactions. In its derivation, the singlet and triplet potentials $V_{S=0}(r)$ and $V_{S=1}(r)$ are simply replaced by their respective scattering lengths. Thus the potentially important effects of magnetic dipoles and hyperfine structure as well as the variation of $\eta$ with the internuclear distance $r$ are neglected, an effect which we will adress below using a CC calculation.

Experimentally, we choose $\ket{i}=\rm Li\ket{2}+\rm Na \ket{1}$, which has the only energetically possible decay channel with equal $M_F=1/2$ to the state $\ket{f}=\rm Li\ket{1}+\rm Na \ket{2}$. The energy gain $E$ is given by the Breit-Rabi formula and by taking also the small magnetic field dependence of $\eta$ into account, one sees that the spin exchange rate is maximized for $B=34$\,G. Therefore, $P_{i \rightarrow f}(B)$ is mapped out for $B=0$\,G to $B=150$\,G and we fit the observed maximum of the loss rate to obtain $P_{i \rightarrow f}(34\,\rm G) = (7.8\pm4.6)\cdot10^{-15}\rm cm^3/\rm s$. With $\eta=0.056$, using eq.\,(\ref{eq:diff}) we finally get $|a_s-a_t| = 5.9^{+1.5}_{-2.1}a_0$, where $a_0$ is the Bohr radius. Due to the simplifications mentioned above, this result can only be considered an estimate, thus we performed a CC calculation with fixed triplet scattering length $a_t=-76a_0$. By varying the singlet scattering length $a_s$, the experimentally observed spin exchange rate is theoretically obtained for $|a_s-a_t|=7.0^{+2.2}_{-3.2} a_0$, in excellent agreement with the value from the approximate formula eq.\,(\ref{eq:diff}).

For all experimental results presented above, the statistical errors in the scattering length values are given by the relative uncertainties in determination of $\Gamma$ ($\sim0.3$), atom numbers $N_{\rm Na}$ and $N_{\rm Li}$ ($\sim0.1$), temperature $T$ ($\sim0.1$) and condensate fraction ($\sim0.2$) \cite{schuster2012phd}. 

\section{Modeling the Feshbach resonances}\label{Theory}

Feshbach resonances occur when molecular states are resonant with two scattering atoms. For Li and Na atoms in the ground state $^2S_{1/2}$ the molecular states are related to the least-bound rovibrational levels of the singlet \Xstate~and triplet \astate~Born-Oppenheimer potentials. Their rovibrational levels are labeled by the vibrational and rotational angular momentum quantum numbers $v$ and $l$ as well as the space-fixed projection $m_l$ of the rotational angular momentum along the magnetic field direction. The basis $|\sigma\rangle=\ket{S M_S  m_{i_A}  m_{i_B}}$ for a pair of atoms $A$ and $B$ describes the remainder of the molecular state, where the fixed atomic quantum numbers $(s_A, i_A, s_B, i_B)$ for electron and nuclear spin are suppressed in the basis vector for brevity. The total electron spin of the molecule is given by $S$. Projections are always defined along the magnetic field direction. Because of the hyperfine interaction it is useful to introduce the coupled basis $\ket{(f_A f_B)f,m_f}$, in which $\vec{f}=\vec{f}_A+\vec{f}_B$, and $m_f$ is its projection. We note that $m_f=m_{f_A}+m_{f_B}=M_S+m_{i_A}+m_{i_B}$. The interaction Hamiltonian conserves the total angular momentum $\vec F=\vec f+\vec l$ at zero magnetic field, and always conserves its projection $M_F=m_f+m_l$.

We focus on ultra-cold collisions, with temperatures well below the $p$-wave centrifugal barrier. Hence, in general only $s$-wave collisions need to be considered, although $p$-wave scattering may be enhanced at $p$-wave resonances. However, $l$=2 molecular states can induce Feshbach resonances in the $s$-wave channels, since distinct angular momentum states are coupled via the weak magnetic dipole spin-spin interaction. This anisotropic interaction does not conserve the quantum numbers $l$ and $m_l$, and gives rise to the selection rules $|l'-l|=0$ and~2, and $|m_l'-m_l|\le 2$. The number of $l=2$ states for a given $M_F$ is much larger than for $l=0$, therefore it gives rise to many more possible $d$- than $s$-wave resonances. Note that the $d$-wave resonances show up as resonances in the $s$-wave scattering length, and therefore in practice are not different from direct $s$-wave Feshbach resonances.

\subsection{Assignment by ABM}\label{ABM}

We have applied the asymptotic bound-state model (ABM) \cite{wille2008eau,tiecke2010abs}, which provides a powerful yet computational light description of the near threshold molecular spectrum, in order to assign the observed Feshbach resonances. By using the binding energies of the rovibrational states and the wavefunction overlap between singlet and triplet states as fit parameters it circumvents the need of detailed knowledge on the molecular potentials and wavefunctions. The ABM is very suitable to guide both experiment and coupled-channel calculations once a few resonances are found. We define $B_0^{\rm ABM}$ as the crossing point of a molecular state and the atomic threshold. In principle, this differs from the Feshbach resonance position because finite coupling between the molecular bound states and the threshold channel shifts the resonance position \cite{tiecke2010abs,chin2010fri}. However, since the observed Feshbach resonances are very narrow, these shifts can be neglected. 

\begin{figure}
\includegraphics[width=8.5cm]{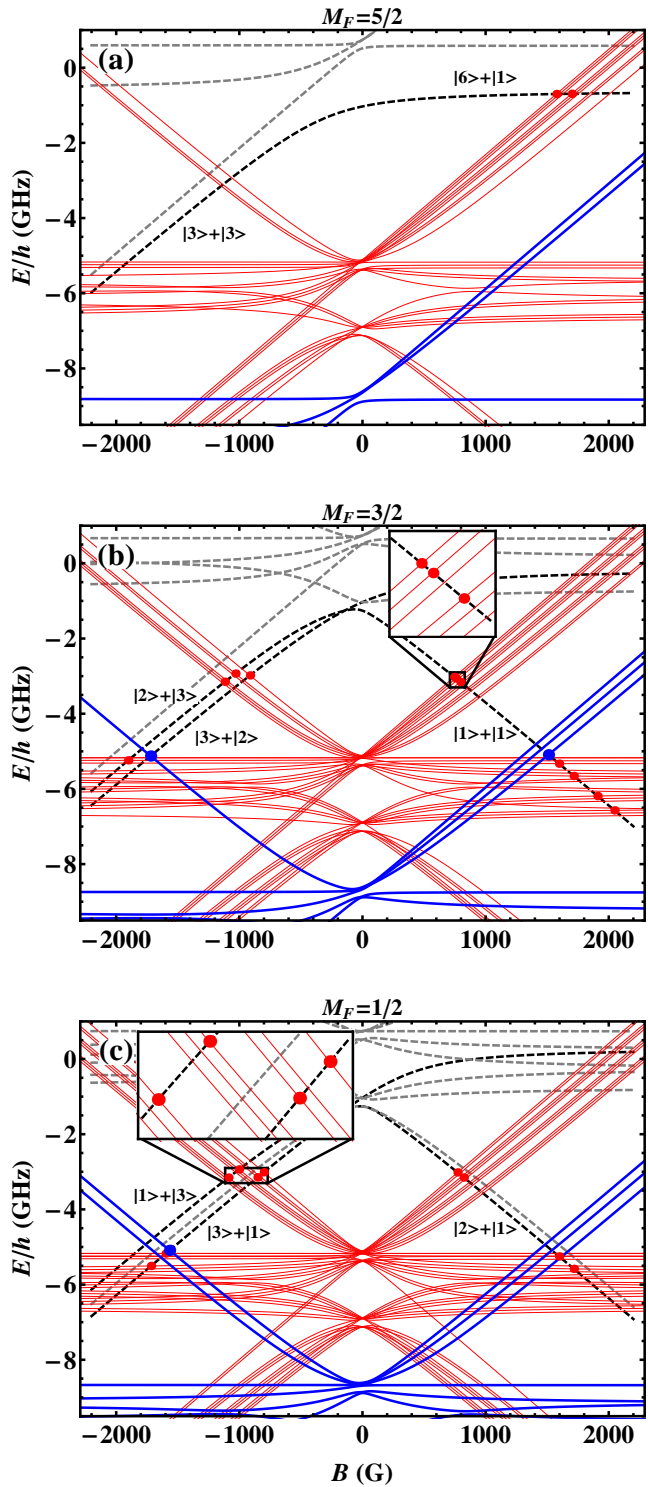}
\caption{(color online) Near-threshold molecular spectra for different $M_F$ from ABM, showing the molecular states $l=0$ (thick blue) and 2 (solid red) states, and the observed $s$- (large blue) and $d$-wave (small red) Feshbach resonances as filled circles. Atomic thresholds are depicted as black and gray dashed lines. The black and labeled ($|\textsuperscript 6{\rm Li}\rangle+|{\rm Na}\rangle$) thresholds are those relevant for the observed Feshbach resonances. The inset in panel (b) shows the resonances in the $|1\rangle+|1\rangle$ channel, which also have been previously observed \cite{stan2004oof}. In (c) a zoom in of four $d$-wave resonances around 900\,G exemplarily highlights the fit quality in the $|1\rangle+|3\rangle$ and $|3\rangle+|1\rangle$ channel. A negative magnetic field corresponds to a sign change in $M_F$.\label{overview}}
\end{figure}

In our attempt to match the near-threshold molecular spectrum to our observed Feshbach resonances, it became clear that the previous assignment by Stan \etal~\cite{stan2004oof} and the predictions of Gacesa \etal~\cite{gacesa2008fri} could not hold. From our measurements of the scattering length, we estimate the energies of the least bound singlet ($S=0$) and triplet ($S=1$) states, $\varepsilon_0^S$. Using the results of Ref.\,\cite{raab2008qff} and the value of the long-range coefficient $C_6$ from Ref.\,\cite{derevianko2001hpc}, we obtain $-10.4$\,GHz\,$< \varepsilon^{S}_0/h < -9.7$\,GHz, with the bound state energy being given with respect to the atomic hyperfine multiplet barycenter. Here, we do not distinguish between triplet and singlet, as the uncertainty in the absolute value of $a_t$ exceeds the small experimentally measured difference $|a_s-a_t|$, an argument which then also applies to the bound state energies. With this result used as parameter in the ABM, the appearance of $s$-wave resonances below 1.5\,kG can be ruled out. Furthermore, assigning the  observed resonances with position $B_0^{\rm exp}<1.5$\,kG as $p$-waves leads to an inconsistent scenario. Identifying the first group of resonances as caused by molecular states with $l=2$, a consistent picture was found, in which all 26 resonances are assigned. The resulting molecular spectra for the three different $M_F$ quantum numbers considered are shown in Fig.\,\ref{overview} as obtained from ABM. Negative magnetic field corresponds to a sign change in $M_F$. The dots represent the observed Feshbach resonances. We note that many $s$- and $d$-wave Feshbach resonances are not observed in the investigated magnetic field range, consistent with the findings of the coupled-channel calculation revealing that their widths are below the detection limit of our apparatus (see section \ref{CC}).

Since our measured resonance spectra contain both $s$- and $d$-wave resonances, we need in principle six ABM parameters: the four binding energies $\varepsilon^{S=0,1}_{l}$ and overlap parameters $\eta_{l}$, where latter can differ for $l=0$ and $l=2$. However, since the corresponding bound states are weakly-bound and thus have a large spatial extent, the rotational splittings $\varepsilon_{\rm rot}^{S}=\varepsilon^{S}_{l=2}-\varepsilon^{S}_{l=0}$ between the $l=0$ and $l=2$ bound states are approximately equal for singlet and triplet states, i.\,e. $\varepsilon_{\rm rot}^{S=0}=\varepsilon_{\rm rot}^{S=1}$. Moreover, we assume that the overlap parameters $\eta_{l}$ for $l=0$ and $l=2$ waves are equal. Since we have mostly resonances of $d$-wave character, we choose the four remaining parameters to be the $d$-state binding energies $\varepsilon^{0}_2$ and $\varepsilon^{1}_2$, the overlap $\eta_2$, and the $d$-wave rotational shift $\varepsilon_{\rm rot}$. We adjust these parameters to fit the molecular spectrum to the observed Feshbach resonances, using a weighted least-square fit procedure. We obtain $\varepsilon^{0}_2/h=-5.949$\,GHz, $\varepsilon^{1}_2/h=-5.851$\,GHz, $\eta_2=0.982$, and $\varepsilon_{\rm rot}/h=3.501$\,GHz. The corresponding Feshbach resonance positions $B_0^{\rm ABM}$ are given in Table\,\ref{FRlist}. We see that the ABM reproduces all but one resonance positions within 1\,G, providing very strong evidence of the correctness of our assignment.

\subsection{Coupled-channel calculation}\label{CC}

Guided by the ABM results, we performed a coupled-channel calculation that includes the full interaction potentials, to obtain a more thorough description of the Feshbach spectrum. To describe the relative motion of two scattering atoms in the coupled system \Xstate+\astate~we use the following Hamiltonian (see~e.\,g.~\cite{stoofsea1988,laue2002mfi}):
\begin{equation}
H = T+ H_{\rm hf}(R) + H_Z+  \sum_{S=0,1} P_S V_S(R) + V_{\rm dip}(\vec R)
\label{ccham}
\end{equation}
where $T$ is the operator for the relative kinetic energy of the atoms,
\[
H_{\rm hf}(R)=\sum_{\alpha=A,B}a_\alpha(R)\vec{s}_\alpha\cdot\vec{\imath}_\alpha/\hbar^2
\] is the $R$-dependent hyperfine-contact interaction of atom A~and B, and
\[ H_Z=\sum_{\alpha=A,B}
(g_{s\alpha}s_{z\alpha}+g_{i\alpha}i_{z\alpha}) \mu_B B/\hbar
\]
is the magnetic Zeeman interaction. $V_S$ are the Born-Oppenheimer potentials for the singlet and triplet states with their projectors $P_S$. Finally, $V_{\rm dip}(\vec R)$ is the weak but non-negligible magnetic dipole spin-spin interaction that includes the second-order spin-orbit interaction. Nevertheless, in the fitting procedure the latter extension turned out to be insignificant for the present data set of Feshbach resonances for LiNa. $V_{\rm dip}(\vec R)$ depends on the orientation of the electron spin with respect to the internuclear axis $\vec R$ as well as the internuclear separation and therefore couples and thus mixes different partial waves $\vec l$. The atomic constants are taken from Ref.\,\cite{arimondo1977edo}, the atomic masses of Li and Na from Ref.~\cite{audi2003}.

The Born-Oppenheimer potentials are introduced as power expansions with a non-linear function of internuclear separation $R$, and are defined in an accompanying paper \cite{tiemann2011}. They are derived  from a fit to all measured Feshbach resonances and recently measured Fourier spectroscopy data of more deeply bound rovibrational levels of the \Xstate~potential as presented in the same paper \cite{tiemann2011}. The fit is split into two parts because of the very different kinds of data. Feshbach resonances depend strongly on hyperfine and Zeeman interaction, whereas deeply bound rovibrational levels from Fourier spectroscopy belong to almost uncoupled singlet and triplet states. Despite the fact that experimentally most Feshbach resonances are detected via three-body loss, the coupled-channel calculation determines their position from the two-body resonance. For the fit of the Feshbach resonances, the slope of the short range branch of both potentials is varied first only for $l=2$, because of the large number of observed $d$-wave resonances.

The $s$-wave resonances are assigned only after a good model for the $d$-wave resonances is obtained. With this result we predict where $s$-wave resonances could occur and found three in our set of observations about 12\,G off the predictions. This tells immediately that the rotational splitting is not yet correctly described by the applied long range function, which is introduced by dispersion coefficients $C_6$, $C_8$ and $C_{10}$ as reported from theory by Ref.\,\cite{derevianko2001hpc,porsev2003arm}. Because we have only few data from the $s$-wave Feshbach resonances, a fit of all three dispersion parameters would be meaningless as it leads to strongly correlated values. We select $C_6$ for variation. From preliminary potentials we find that the outer turning points of the relevant rovibrational levels will lie around 32$a_0$ and thus the influence of the higher order terms is much less than that of $C_6$. The fits yield~a~$C_6$ value which is 2.2\% smaller than the theoretically predicted one \cite{derevianko2001hpc}. This is roughly twice the uncertainty expected from that calculation. We do not interpret our finding as a deviation because of the correlation to the higher order terms which we keep constant.

As described in Ref.\,\cite{tiemann2011} the fit routine runs iteratively through the above mentioned two parts. With the result of the Feshbach fit, the Born-Oppenheimer potentials can predict the position of the uncoupled least bound singlet and triplet levels. They are used in the second part of the fit together with all deeply bound rovibrational levels to obtain potentials getting an improved description of Feshbach and spectroscopy data. Then the separated Feshbach fit is started again and later with this new result the combined spectroscopy fit. Only few iterations are needed for convergence. With the coupled-channel calculation we can select from the many possible Feshbach resonances (see Fig.\,\ref{overview}) the ones which are broad enough to become detectable in the present experimental setup, i.e. with $\Delta\gtrsim0.02$\,mG. This selection was confirmed by the additional observations of the resonances at 800.9\,G and 1700.4\,G, which proves the predictive power of our scenario. In total, all theoretically predicted $M_S=1$ resonances with $\Delta>0.02$\,mG were experimentally observed, whereas for $M_S=0$ no further systematics were done after the successful resonance assignment.

\begin{table}
\caption{Results of the coupled-channel calculation on the $\textsuperscript 6$Li+Na singlet \Xstate~and triplet \astate~background scattering lengths and the binding energies of the least bound states.}
\label{scatteringlengthlist}
\begin{ruledtabular}
\begin{tabular}{c  c c }
                        &  $S=0$        &   $S=1$      \\
\hline\\[-8pt]
$a$($a_0$)              &  -73(8)       & -76(5)        \\
$\varepsilon^{S}_0/h$(GHz) &  \!\!\!-9.3838(50)  & -9.35335(50)  \\
$\varepsilon^{S}_2/h$(GHz) & -5.95634(40)  & -5.85180(30)   \\
\end{tabular}
\end{ruledtabular}
\end{table}

The results of the coupled-channel calculation in terms of the Feshbach resonance position $B_0^{\rm CC}$ and width~$\Delta$, defined as the separation of field positions between the peak and the zero crossing of the scattering length, are given in Table~\ref{FRlist}. All resonances are modeled well within the experimental uncertainty. The most striking result is that the $s$-wave resonance widths do not exceed 0.1\,mG, while the $d$-wave widths can reach 10\,mG. For the two resonances of asymptote $|6\rangle+|1\rangle$ no widths are given, because the resonances are significantly influenced by inelastic decay channels. Thus the calculated functional form of the scattering lenght is not described by the normally assumed mathematical form as it does not necessarily show a zero crossing at all. This might also explain why in the ABM fit these two resonances show the largest deviation. 

From the improved $V_{S=0}$ and $V_{S=1}$ potentials the background scattering lengths can be extracted, as well as the binding energies of the least bound states, whose values are given in Table~\ref{scatteringlengthlist}. We notice that both ABM triplet values deviate less than 1\,MHz from those obtained from the full interaction potentials, while the $d$-state singlet value agrees within 7\,MHz. The scattering length values, which have been derived from the results of the coupled-channel calculation with the Feshbach resonance spectrum as input, are in perfect agreement with the independently obtained experimental results $a_t=-69(13)a_0$ and $|a_t-a_s|=5.9^{+1.5}_{-2.1}a_0$. 

It is important to note that the singlet and triplet scattering lengths are nearly equal, which is intimately linked to the narrow widths of the Feshbach resonances arising due to a combination of factors: First, the singlet and triplet interaction potentials are far from being resonant, indicated by their scattering lengths which are of the same order as the van der Waals length $r_0=\frac{1}{2}(2\mu C_6/\hbar^2)^{1/4} = 35a_0$, where $\mu$ denotes the reduced mass. Second, the effective coupling between singlet and triplet states is very small, indicated by $|a_s-a_t|\ll r_0$. A similar situation can be found in homonuclear $^{87}$Rb \cite{marte2002fri}, where the narrowness of the Feshbach resonances and the smallness of the loss rates can be traced back to non-resonant and similar values of $a_s$ and $a_t$ \cite{kokkelmans1997roc}.  

\section{$^7$Li+Na}\label{7LiNa}

\begin{figure}
\includegraphics[width=8.5cm]{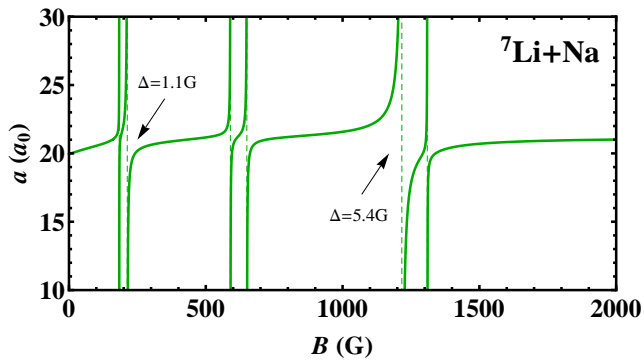}
\caption{Scattering length of $^7$Li+Na in the energetically lowest spin channel, derived from the interaction potentials. The spectrum shows much broader $s$-wave Feshbach resonances than found in $\textsuperscript 6$Li+Na, most notably a 5.4\,G broad resonance around 1220\,G.
\label{scat_7Li}}
\end{figure}

The lack of a broad $s$-wave Feshbach resonance to tune the interaction in ultracold $\textsuperscript 6$Li+Na, has motivated us to consider the isotopologue Bose-Bose $^7$Li+Na mixture. From the improved interaction potentials obtained for $\textsuperscript 6$Li+Na, predictions of scattering lengths and possible Feshbach resonances in $^7$Li+Na can be made within the Born-Oppenheimer approximation. Fig.\,\ref{scat_7Li} shows the predicted scattering length of the energetically lowest spin channel of $^7$Li+Na as an example. Here much broader Feshbach resonances are present than in $\textsuperscript 6$Li+Na, including a 5.4\,G broad resonance around 1220\,G. We estimate that the accuracy of the Feshbach resonance prediction to be better than 50\,G, keeping the overall structure and width of the resonances as given in Fig.\,\ref{scat_7Li}. The main uncertainty of these predictions originates from the insufficiently characterized well depth of the triplet potential \cite{tiemann2011}. The singlet and triplet scattering lengths are predicted to be around $5a_0$ and $21a_0$, respectively.

\section{Conclusion and Outlook}\label{concl}

We have observed 26 interspecies Feshbach resonances in ultracold $\textsuperscript 6$Li+Na, consisting of 3 $s$- and 23 $d$-wave resonances. The broadest resonance has a width of 10\,mG, which practically limits the tunability of the interaction by a magnetic field. On basis of our extensive Feshbach spectra and new Fourier spectroscopy data \cite{tiemann2011} the interaction potentials have been improved, leading to the scattering lengths $a_s=-73(8)a_0$ and $a_t=-76(5)a_0$. Within the mutual uncertainties, they show very good agreement with our experimental findings both in sign and magnitude, where the latter was obtained via an oscillation measurement. The relatively large absolute value of the triplet scattering length explains the efficient sympathetic cooling observed in Ref.\,\cite{hadzibabic2002tsm,hadzibabic2003fii}. Moreover, the experimentally measured long lifetimes of different spin mixtures are explained by the small difference of $a_s$ and $a_t$, which we also confirm quantitatively.

The scattering lengths determined in this study show still fairly large errors (see table \ref{scatteringlengthlist}), which result from the uncertainty in the long range function or more explicitly from the uncertainty in $C_6$ due to high correlation between the dispersion coefficients. One could gain information by measuring $s$-wave resonances which have significant singlet character, as the ones reported here are of strong triplet character only. Examples of such resonances are expected to be at 2981(3)\,G for the entrance channels $|1\rangle+|1\rangle$ and $|2\rangle+|1\rangle$ and at 3137(3)\,G for $|1\rangle+|1\rangle$ and will show widths around 10\,mG thus similar to the broadest observed $d$-wave resonances. But such high fields are not reachable in the present experimental setup. The lack of data on the triplet potential could be overcome by two-color photo-association experiments ending in deeply bound triplet levels. For such an experiment, Feshbach molecules could be prepared via the broadest $d$-wave resonances which have strong triplet character. This association process should still be possible for resonances having a width larger than 1\,mG \cite{mark2005eco,mark2007sou}.

\begin{acknowledgments}

We thank Mathias Neidig and Andrea Bergschneider for setting up part of the Li laser setup and Axel G\"{o}rlitz for useful comments on our magnetic trap. We acknowledge support from the Heidelberg Center of Quantum Dynamics and the ExtreMe Matter Institute. T.\,S.\, and R.\,S.\, acknowledge support from a Klaus-Tschira Scholarship and a scholarship of the Landesgraduiertenf\"orderung Baden-W\"urttemberg, respectively. E.\,T.\, thanks the cluster of excellence ``QUEST" for support and the Minister of Science and Culture of Lower Saxony for providing a Niedersachsenprofessur. S.\,K.\, acknowledges financial support from the Netherlands Organization for Scientific Research (NWO) via a ``VIDI" grant.

\end{acknowledgments}

\end{document}